# First Neutron Spectrometry Measurement at the HL-2A Tokamak[*]

YUAN Xi(袁熙)[1], ZHANG Xing (张兴)[1], XIE Xufei (谢旭飞)[1],

CHEN Zhongjing (陈忠靖)[1], PENG Xingyu(彭星宇)[1], FAN Tieshuan (樊铁栓)[1,1)],

CHEN Jinxiang (陈金象)[1], LI Xiangqing(李湘庆)[1], YUAN Guoliang (袁国梁)[2],

YANG Jinwei（杨进蔚）[2], YANG Qingwei（杨青魏）[2]

[1] State Key Laboratory of Nuclear Physics and Technology,

Peking University, Beijing 100871, China

[2] Southwestern Institute of Physics, Chengdu 610041, China

Abstract：A compact neutron spectrometer based on the liquid scintillator is presented for the neutron energy spectrum measurement at the HL-2A tokamak. The spectrometer has been well characterized and a fast digital pulse shape discrimination software has been developed using the charge comparison method. A digitizer data acquisition system with the maximum frequency of 1 MHz can work under the high count rate environment at HL-2A. Specific radiation shielding and magnetic shielding for the spectrometer has been designed for the neutron spectrum measurement at the HL-2A tokamak. For the analysis of the pulse height spectrum, dedicated numerical simulation utilizing NUBEAM combining with GENESIS has been made to obtain the neutron energy spectrum, following which the transportation process from the plasma to the detector has been evaluated with Monte Carlo calculations. The distorted neutron energy spectrum has been folded with response matrix of the liquid scintillation spectrometer, and good consistency has been found between the simulated pulse height spectra and the measured ones. This neutron spectrometer based on digital acquisition system could be well adopted for the investigation of the auxiliary heating behavior and the fast ion related phenomenon on the different tokamak devices.

[*] Supported by the state Key Development Program for Basic Research of China (Nos. 2013GB106004, 2012GB101003 and 2008CB717803) , the National Natural Science Foundation of China (No. 91226102 ) and the National Science and Technology support program (No. 2011BAI02B01).

[1] E-mail: tsfan@pku.edu.cn

# 1 Introduction

Neutron diagnostics have received growing interest in last decades due to their role of assessing certain plasma parameters, such as fusion power, ion temperature, fast ion energy and their spatial distributions in the plasma core[1,2]. The Magneto-Hydro-Dynamic (MHD) induced fast-ion losses phenomenon has been both experimentally observed and theoretically predicted [3,4] in different tokamak experiments. Similar MHD activities, such as sawbone, ion-fishbone and long-lived bone have been observed at HL-2A [5,6]. To reveal the fast-ion related phenomena, it is important to measure the fast ion distribution. Neutron yield and energy spectrum are well correlated with the fast-ion distribution, for which it is widely used for the assessment of the fast-ion behavior [7,8,9]. At HL-2A, neutral beam injection (NBI) heating is an important source of fast ions. As for the relatively low ion temperature, beam-thermal neutrons are dominant among all the neutron species. Therefore, it is of advantage to evaluate fast ions with neutron measurements. A liquid scintillator spectrometer is generally employed as for its high efficiency, good neutron/gamma-ray discrimination ability and compact size which could be located near to the device. Now many neutron spectrometers based on liquid scintillator have been installed on tokamak devices, such as JET [10,11], AUG [12], FTU [13], JT-60U [14], MAST [15] and EAST [16].

In this work, a compact neutron spectrometer (CNS) based on the liquid scintillator EJ301 has been developed and operated for the neutron spectrum measurements at the HL-2Atokamak. The spectrometer has been well characterized and a fast digital pulse shape discrimination software has been developed using the charge comparison method. For the analysis of the pulse height spectrum, dedicated numerical simulation utilizing NUBEAM [17] combining with GENESIS [18,19] has been made to obtain the neutron energy spectrum, following which the transportation process from the plasma to the detector has been evaluated with Monte Carlo calculation code MCNP [20]. Good consistency has been found between the simulated pulse height spectra and the measured ones.

This paper is organized as follows: the experimental setup, calibration results of CNS and digital pulsed shape discrimination techniques are presented in section 2. The experimental measurements and Monte Carlo analysis of CNS at HL-2A are given in section 3. Then section 4 describes the measured results of CNS, followed by the conclusions in section 5.

## 2 Energy Calibration and digital pulsed shape discrimination of the liquid scintillator neutron spectrometer

The experimental arrangement of the compact neutron spectrometer is schematically shown in Fig.1. The core cell of the CNS was a cylindrical liquid scintillator EJ301. The liquid scintillator had an active volume of near 100 $cm^3$ (2 inch in diameter, 2 inch in length), and the thickness of the liquid scintillator was chosen to be 2 times larger than the mean range of 2.45MeV protons in the sensitive material. The scintillator was optically coupled to a 2 inch Hamamatsu R329-02 photomultiplier tube (PMT) directly, and the PMT was coupled to an ORTEC 265A voltage divider and magnetic shielding was composed by an additional 0.8 mm $\mu$-metal shield. The anode signals of CNS were sent into a high-speed digitizer of model Agilent U1065A through a 2 m long coaxial-cable. Agilent U1065A was a quad-channel 10-bit digitizer with sampling rate up to 8 Gsamples/s, and the full scale range was from 50 mV to 5 V. Raw data were transferred to a PC over an Ethernet connection using UDP/IP.

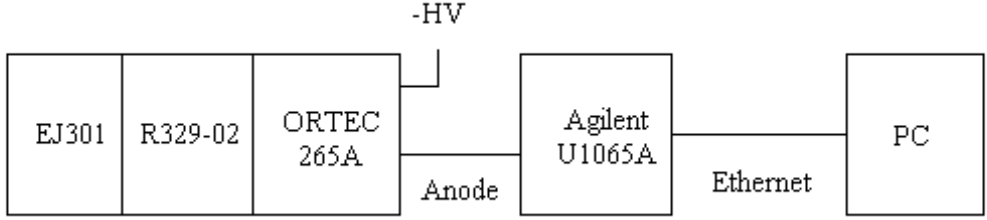

Fig. 1 Schematic diagram of the compact neutron spectrometer.

The pulse height spectrum (PHS) of CNS could reflect the light output for secondary particles such as electrons and protons, and it could be considered as the photon and neutron spectrum folded with the response matrix. For the neutron spectrum measurement, it is important to get the response matrix specifically for this spectrometer. Therefore, calibration experiments have been established with gamma-ray and neutron sources.

Energy calibration was performed with gamma-ray sources ($^{137}$Cs, $^{54}$Mn, $^{22}$Na) to establish the relationship between pulse heights $Q_\alpha$ and electron equivalent energy $E_{ee}$, expressed by

$$Q_\alpha = G(E_{ee} - E_0) \qquad (1)$$

Where $Q_\alpha$ is defined as the charge integral of the pulse in a 130 ns time range, G is the calibration factor and $E_0$ is the nonlinearity from the quenching effects and other factors. G and $E_0$ were obtained by iterative fitting between experimental spectra and theoretical spectra calculated by the Monte Carlo code GRESP [21], which was developed by the Physikalisch-Technische Bundesanstalt (PTB). Three experimental configurations of the detector and digitizer (high voltages and full ranges) were used to cover the neutron energy range from 1 MeV to 17 MeV, and the energy calibration was completed for each configuration.

The relationship betweenthe light output function of recoil protons and the electron equivalent energy is strongly nonlinear [22], and this relationship is different for each spectrometer. 15 quasi-monoenergetic neutron fields were generated by nuclear reactions via the 4.5MV Van der Graff Accelerator at Peking University, including Li$^7$(p, n)Be$^7$ (1.14 MeV and 1.38 MeV), T(p, n)$^3$He (1.60 MeV, 1.90 MeV,

2.14 MeV and 2.40 MeV), D(d, n)$^3$He (2.95 MeV, 3.95 MeV, 4.11 MeV, 5.21 MeV and 6.0 MeV) and T(d, n)$^4$He (14.1 MeV, 15.1 MeV, 16.1 MeV and 17.7 MeV). The quasi-monoenergetic neutrons with specific energies were then selected by time-of-flight window methods, which could reduce the ratio of scattered neutrons remarkably. The injected neutron energy and broadenings were calculated by the code TARGET [23], which was also developed by the PTB, and they were adopted as the input parameters of the Monte Carlo code NRESP7 [24]. Iterative calculations and fittings were established to obtain the nonlinear light output function and pulse height resolution $dL/L$. The latter can be expressed as:

$$dL/L = \sqrt{7.5^2 + 8.08^2/L + 0.2^2/L^2} \quad (2)$$

The comparison between experimental and calculated pulse height spectra of $^{54}$Mn and three quasi-monoenergetic neutron fields are shown in Fig. 2.

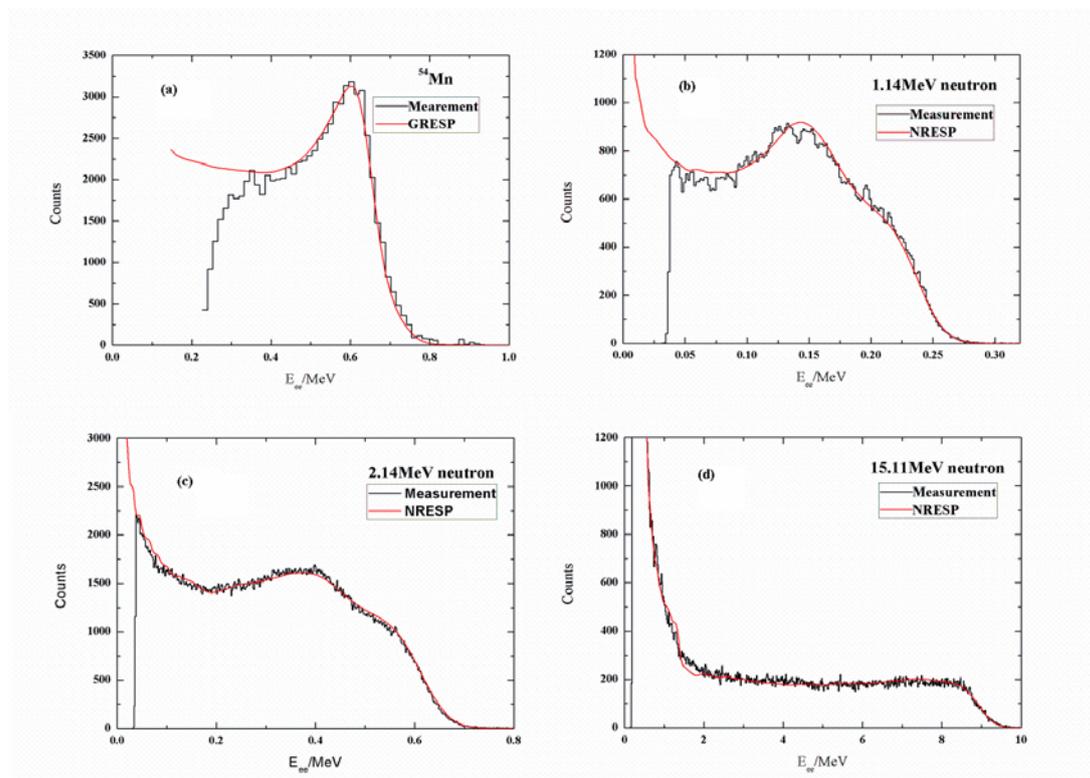

Fig. 2. Comparisons of calculated spectra to measured spectra. The black histograms are the measured spectra. The red lines are the calculated spectra. (a) The measured

andcalculated results from the code GRESP for $^{54}$Mn gamma source, and the measured and calculated results from the code NRESP for 1.14MeV (b), 2.14MeV (c), 15.11MeV (d) quasi-monoenergetic neutron sources, respectively.

The radiation field around the tokamak device is a mixed neutron/gamma-ray field. Sometimes the photon flux is even higher than the neutron flux. These photons are produced by the interactions of runaway electrons with the plasma chamber wall and inelastic scattering and capture reactions of neutrons with the tokamaka device. Therefore, it is necessary to develop a good discrimination technology for the liquid scintillator as it is also sensitive to photons. And the total maximum count rate of the detector must be high enough, and then it could leave enough space for the neutron counting. A new digital acquisition and discrimination system has been developed for these two objectives.

The shapes of the light pulses generated by neutrons and photons in EJ301 liquid scintillator have such difference that could be used for the discrimination between neutron and photon events. The light pulses in the scintillator are the combinations of the similar fast scintillation component (∼ 4-5 ns) and different slow scintillation component (∼100-200 ns). The DPSD used in this work was a typical charge comparison method [25]. Each pulse was integrated for two different time lengths as charge values $Q_{long}$ and $Q_{short}$, denoting the total and fast components, respectively. With the same $Q_{short}$ value, the corresponding $Q_{long}$ value for neutrons is larger than for gamma-rays. As for this EJ301 CNS, the time lengths for the short and long time window were optimized to be 15 ns and 130 ns, respectively. The DPSD results of 2.40 MeV quasi-monoenergetic neutron sources from the T(p, n)$^3$He reaction is shown in Fig. 3. Other DPSD methods based on both time domain analysis and frequency domain analysis have been developed [26,27].

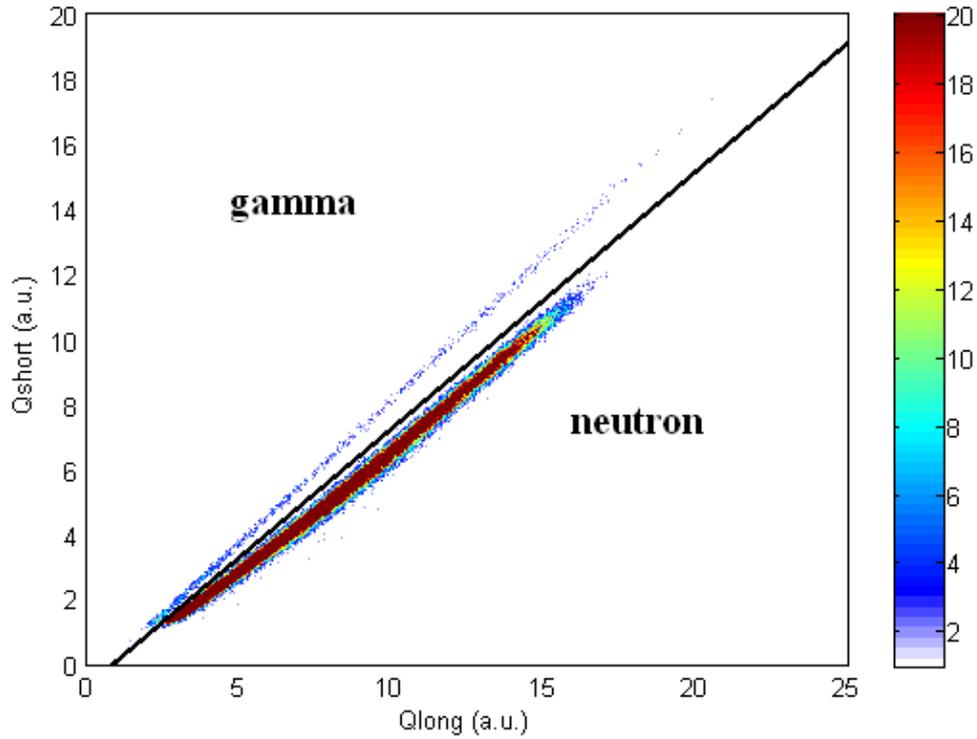

Fig. 3. Plot of $Q_{long}$ versus $Q_{short}$ for 2.40 MeV neutron sources. The straight line which is used to separate neutron events from photon events is shown in the figure.

## 3 Experimental measurements and Monte Carlo analysis at HL-2A

The main plasma discharge parameters of the HL-2A Tokamak are nominally as follows: major radius $R = 1.65$ m, minor radius $a = 0.4$m, $I_P$= 400 kA, $B_T$= 2.65 T, and $n_{e0} \sim 6.0 \times 10^{19}$ m$^{-3}$. During plasma discharges, neutrons are generally produced by the reaction between fusion deuteron fuel ions, while most of neutrons are from the beam-thermal component under NBI which can generate numerous fast fuel ions. Therefore, the neutron energy distribution could be related to the fast ion energy distribution, and it is possible to assess the behavior of fast ions via neutron measurement.

For the application of the CNS at HL-2A, specific shielding has been developed for the radiation and magnetic shielding. The CNS spectrometer was shielded from scattered neutrons by a 40 cm pure polyethylene in the front region. To reduce the weight and size of the shielding, the thickness of polyethylene at the back and side regions were 15 cm, where the fluence of scattered neutrons and photons were low. However, polyethylene could produce 2.226 MeV gamma-rays from the neutron capture reaction, for which a close lead shield of more than 20 cm thickness was located between the polyethylene and CNS. A 1.3 cm DT4C soft iron cylinder between the lead shield and a 2 mm permalloy around the spectrometer were employed to shield the detector from the HL-2A magnetic fields, which was less than 0.02T at the CNS location. The measured spectra of the $^{137}$Cs time by time showed that the gain of detector was stable during discharges. Considering the spatial layout, the shielding was finally located at the mid-plane of HL-2A, and behind the soft X-rays system in the radial direction. The distance between the detector and the nearest plasma core was 3 m, and the sight line of the spectrometer could cover the entire small radius. A photo of shielding in the experimental hall is shown in Fig.4.

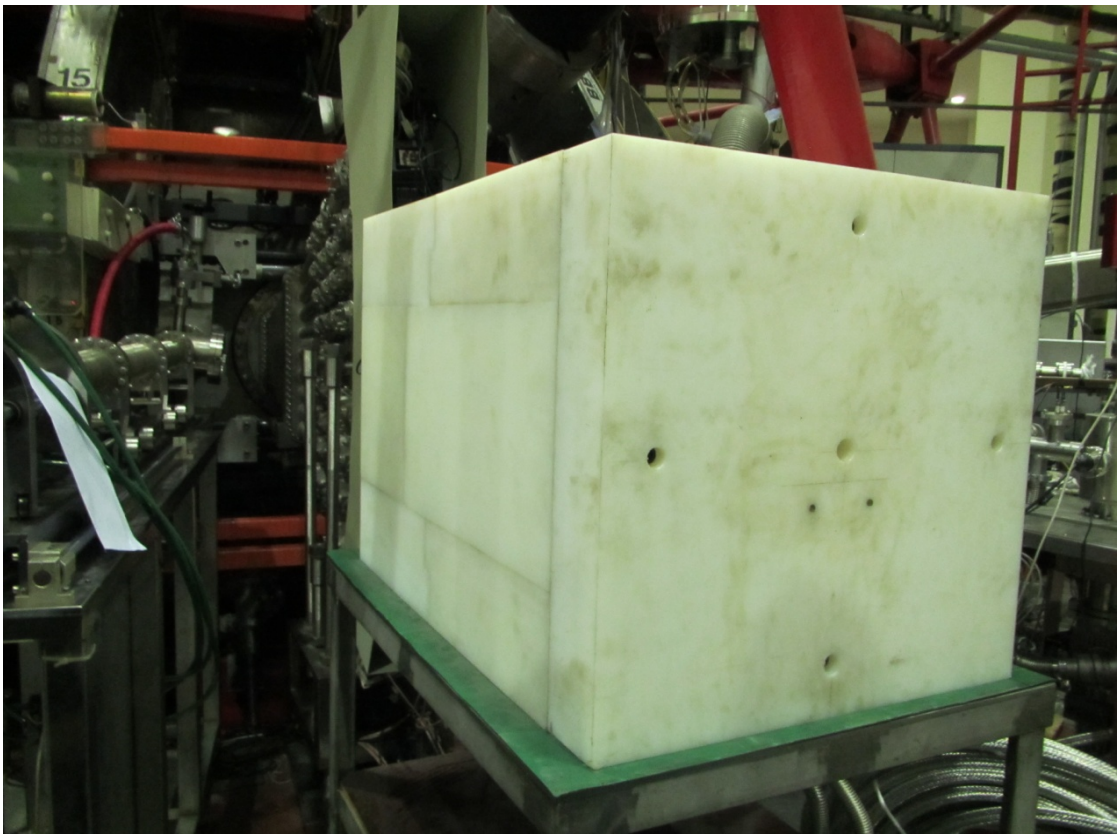

Fig. 4. Scene photo of CNS shielding around HL-2A tokamak.

The picture in Fig. 4 shows the complex environment around the CNS，and it would give rise to the neutron spectrum variations from the plasma to the detector, especially for the components in the line of sight. A 3-D MCNP model was built to simulatethe softening process of the energy spectrum. This model included the main parts of the tokamak, the experimental hall, the faced window, the soft X-ray system and the whole neutron spectrometer. The 2-D schematic diagrams of MCNP model are shown in Fig.5.

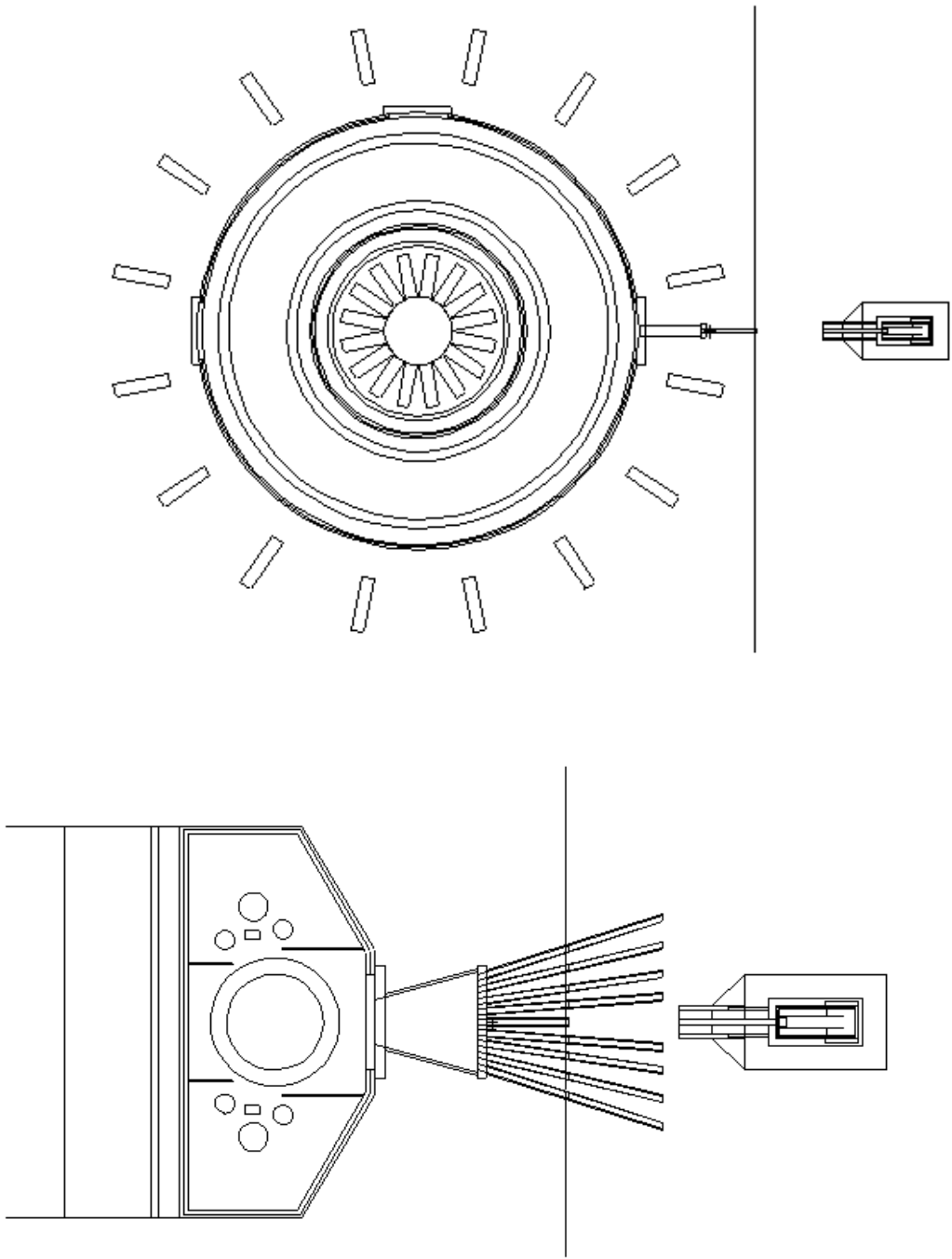

Fig.5. The top plot shows the 2-D schematic diagram of the whole tokamak and shielding at mid-plane, the bottom plot shows the 2-D schematic diagram of the window faced to the shielding.

The emission neutron spectrum was calculated by the Code NUBEAM and the Monte Carlo code GENESIS. NUBEAM simulations were performed for the magnetic equilibrium of the HL-2A shot 20016. The temperature and density profiles, as a function of the normalized toroidal flux surface coordinate, are assumed to be given by

$$T(\rho) = [T(0) - T(1)](1-\rho^2) + T(1) \quad (3)$$

$$n(\rho) = [n(0) - n(1)](1-\rho^2)^{0.5} + n(1) \quad (4)$$

where $T_e(0)$, $T_i(0)$ and $n(0)$ are the temperature and density in the plasma core, and they were set to 1 keV, 0.8keV and $2\times10^{19}$ m$^{-3}$, respectively, $n(1)=0.1\times n(0)$ and $T(1)=0.1\times T(0)$ were further assumed. The plasma dilution was represented by an effective charge number $Z_{eff}=2.5$. Both neutral beam lines had the injection energy of 40 keV and the total power was assumed to be operated at a steady state of 1 MW. The output of the NUBEAM calculation consisted of energy distributions of the beam ions $F_{NBI}$ which were used as the input for calculations of the DD neutron spectrum with the code GENESIS. The code GENESIS, which is based on classical kinematics, is able to calculate the energy spectrum of neutron and gamma-rays emitted from several nuclear reactions in the plasmas, starting from the reactant energy distributions, reaction cross sections and plasma profiles. The neutron spectrum generated in the plasma and the neutron spectrum after transmission are compared in Fig.6. In order to increase the calculation efficiency, the cut energy for neutrons was set to 0.8keV, which was coordinated with the lower threshold of measurement.

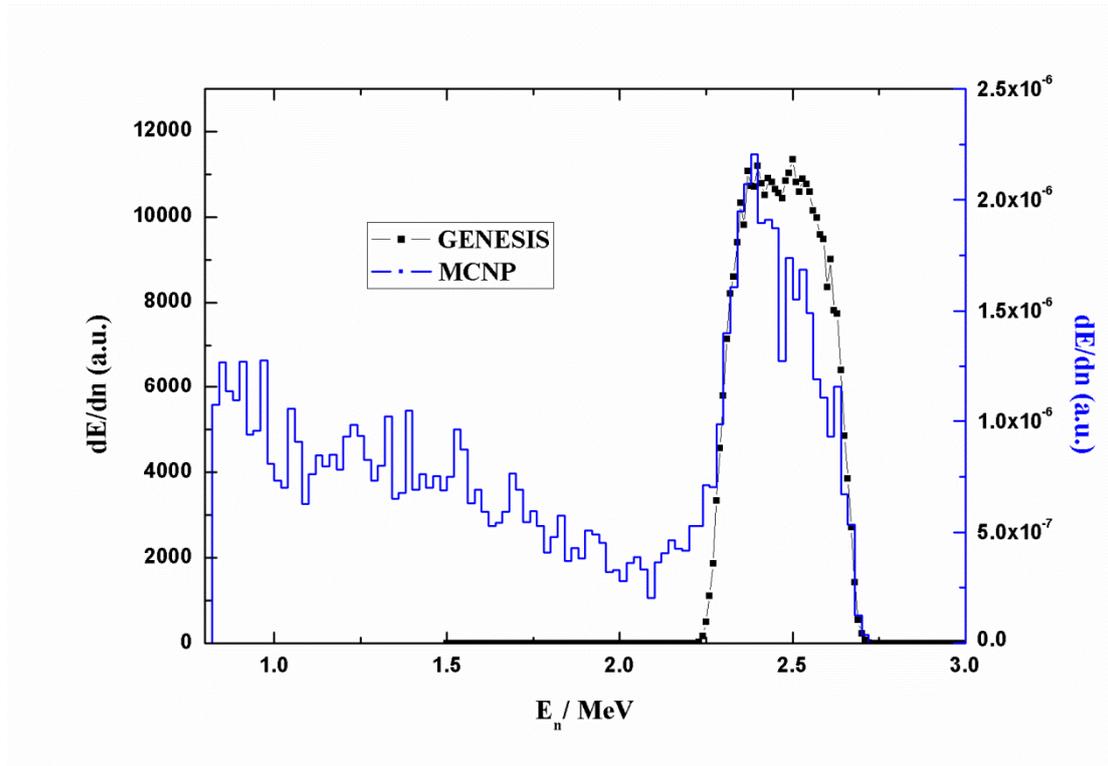

Fig.6. Comparison the input neutron spectrumfrom the code GENESIS (black point) with the output neutron spectrum from the code MCNP (blue step line).

## 4 Experimental results

In some large tokamak devices such as JET and ITER, neutron spectrometers have to acquire and process mass data in a short time (mostly less than 10 s). At HL-2A, the neutron yield is less than $10^8$/s without NBI auxiliary heating and the neutron yield will explosively increase more than $10^4$ times with NBI injections. The maximum count rate for analogue electronic modules isusually less than 100 kHz, which could not meet the need of fusion neutron measurements. For the Agilent U1065A digitizer, there are two working mode. One is the trigger mode and the other is the continuous mode. For the trigger mode, the dead time for each pulse is 1.8 $\mu$s, which means that with the 200 ns record length the maximum count rate is below 500 kHz. For DPSD, the time resolution is set to 1 or 2 ns because of the 4-5 ns rise time of liquid scintillator signals. Therefore, for the largest 2 GB RAM, the record time is 1-2 s. The time interval of NBI is about 500 ms, and without NBI the neutron flux decreases dramatically. In this case, flash ADC was set to 1 Gsamples/s sampling and it began to acquire data 100 ms after the external trigger signals from the clock of HL-2A data acquiring system. The advantage of continuous mode was that digitizer was no longer the limit of maximum count rate and the data transmission speed no longer need to be considered. Instead, more work must focus on the DPSD for the pile-up signals and the strike of PMT for large current. When the CNS was located outside of the shielding, the PMT tube did not work after near 50 ms of NBI heating. The DPSD result for discharge #20016 is shown in Fig.7. Note that the gamma energy is much higher than the neutron energy, which can even reach 8 MeV.

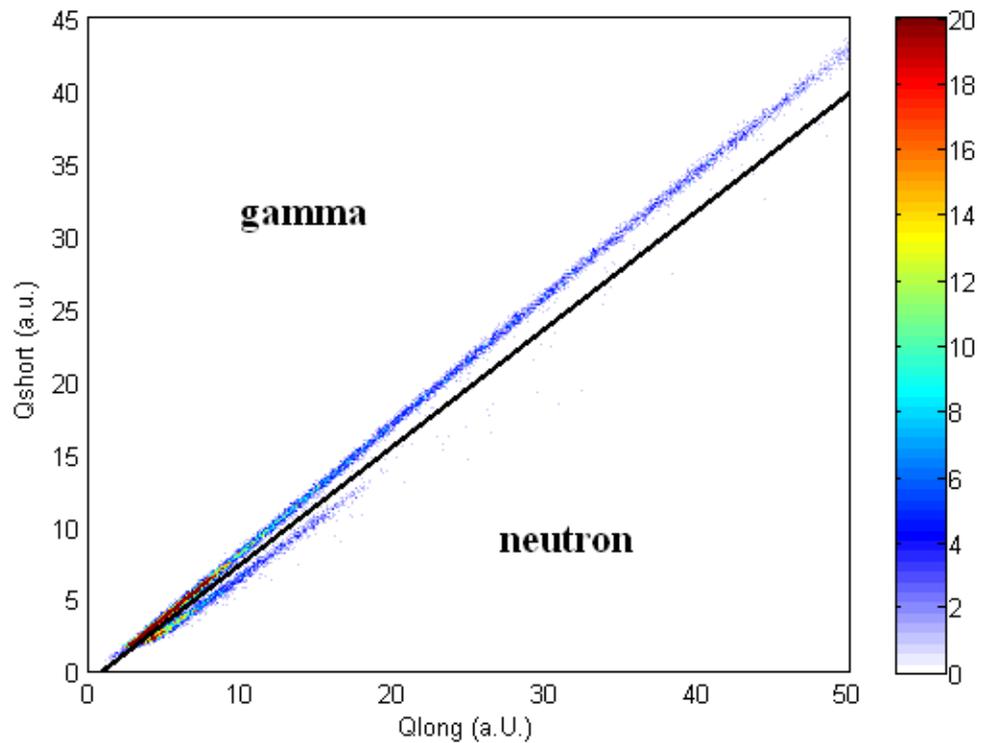

Fig.7. Plot of $Q_{long}$ versus $Q_{short}$ for HL-2A discharge #20016. The straight line which is used to divide neutron events from photons events is indicated in the figure.

The time traces of plasma current $I_p$, electron density $n_e$, NBI power, hard X-ray, neutron fluence monitored by a $^{235}U$ fission chamber and the neutron, gamma-ray fluence from the CNS are demonstrated in Fig.8. Note that the gamma flux with higher energy is at least three times more than the neutron flux and the maximum ratio is 10 times in the experiment scenarios. The time trace of the neutron fluence monitored by the CNS is similar to the one monitored by the fission chamber, which is directly related NBI power. The total count rate of the CNS is near 200k Hz.

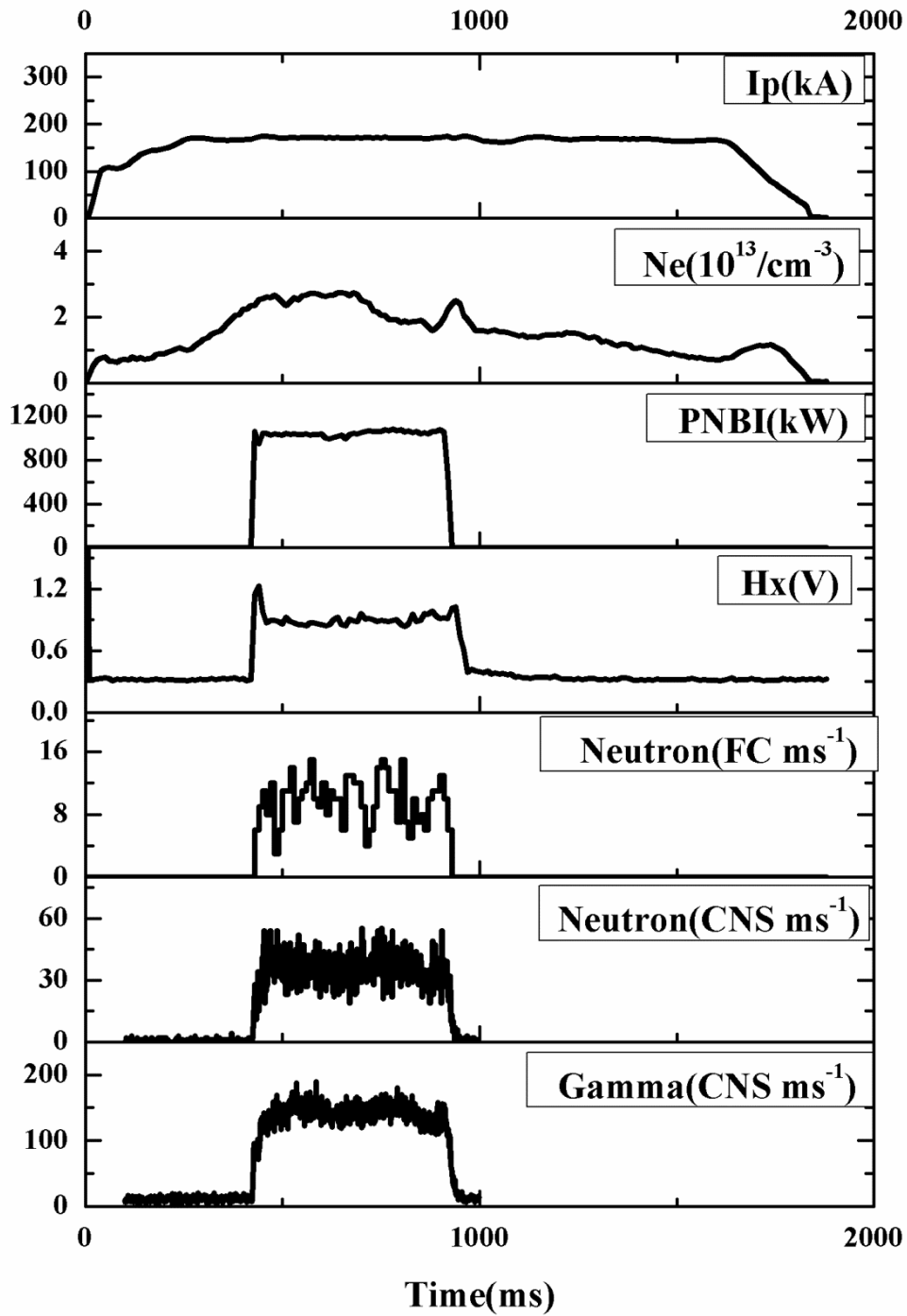

Fig.8. Time traces of plasma current, electron density, NBI power, hard X-ray, neutron and gamma-ray count rate during HL-2A plasma discharge #20016.

In order to increase the statistics of the measured spectra, pulse height spectra of 9 similar discharges (#20016-20025 except #20021) were summed. The neutron spectrum detected at the CNS was the same with the result shown in Fig.6. The theoretical pulse height spectrum was obtained by folding the neutron energy spectrum with the detector response matrix, and it is generally consistent with the experimental results, as illustrated in Fig. 9. It also indicates that neutron emission is dominated by the beam-plasma reactions.

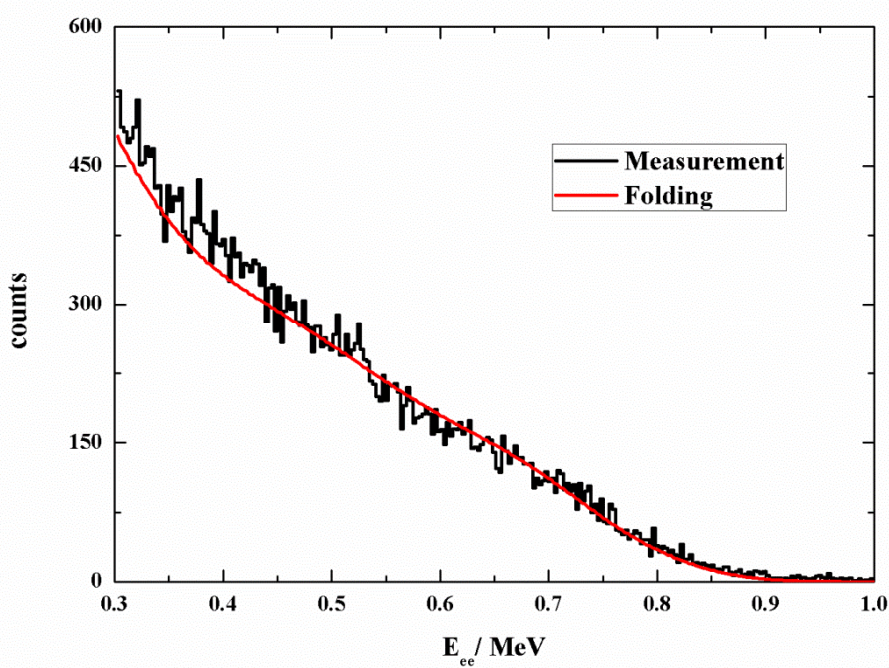

Fig.9. Comparison between the measured pulse height spectra of 9 discharges (#20016-20025 except #20021, black step line) and calculated folding pulse height spectra (red line).

## 5 Conclusions

A compact neutron spectrometer was developed and built at HL-2A for the measurement of neutron emission spectrum. This spectrometer worked well at the complex environment and high photon and neutron scattering background, and the maximum count rate was up to 1MHz. For the NBI heating at HL-2A, beam-thermal neutron component was calculated by the code NEUBEAM and GENESIS, and detailed modeling of neutron transmission from plasma core to detector has been made using the MCNP code. The experimental pulse height spectrum could be well fit with the theoretical one. Future work will be focused on the analysis of neutron spectra in different heating scenarios.

**Acknowledgement**

The authors are very grateful to the HL-2A operation Team for their help during the experiment on HL-2A. The authors are also grateful to Prof. R. Nolte and Dr. M. Reginatto of PTB for the help in using the GRESP, NRESP and TARGET codes. The authors are grateful to Prof. Jianyong Wang of Peking University for the accelerator operation.